\documentclass[12pt]{article}
\usepackage{epsf}
\begin{document} YITP-01-36
\hspace{10cm}
\today
\\
\vspace{3cm}
\thispagestyle{empty}
\begin{center} {\LARGE    Local Casimir energy for solitons
    }
\\  \vspace{1cm} {Alfred Scharff Goldhaber\footnote{e-mail:
goldhab@insti.physics.sunysb.edu},  Andrei Litvintsev\footnote{e-mail:
litvint@insti.physics.sunysb.edu}  and Peter van  
Nieuwenhuizen\footnote{e-mail:
vannieu@insti.physics.sunysb.edu}\\
  { \it C.N.Yang Institute for Theoretical Physics, State University of  
New York},
\\ {\it Stony Brook, NY 11794 } }

\abstract{Direct calculation of the one-loop contributions to the
energy density of bosonic and supersymmetric $\phi^4$ kinks
exhibits:  (1) {\it Local mode regularization}.  Requiring the
mode density in the kink and the trivial sectors to be equal at
each point in space yields the anomalous part of the energy
density.  (2)  {\it Phase space factorization}. A striking
position-momentum factorization for reflectionless potentials
gives the non-anomalous energy density a simple relation to that
for the bound state.   For the supersymmetric kink, our expression
for the energy density (both the anomalous and non-anomalous
parts) agrees with the published central charge density, whose
anomalous part we also compute directly by point-splitting
regularization.  Finally we show that, for a scalar field with arbitrary
scalar background potential in one space dimension, point-splitting
regularization implies local mode regularization of the Casimir energy
density. }
\end{center}
\newpage

{\it Introduction.---}Quantum corrections to solitons were of great  
interest in the
   1970's and 1980's, and again in the last few years, due to the  
present  activity in
quantum field theories with dualities between extended objects and  
pointlike objects.
Dashen, Hasslacher, and Neveu
\cite{dashen}, in a 1974 article that has become a classic,  computed  
the one-loop
corrections
  to the mass of the bosonic kink in
$\phi^4$ field theory and to the bosonic soliton in sine-Gordon theory.  
  For the
latter, there exist exact analytical methods associated with the  
complete
integrability of the system, authenticating the perturbative  
calculation.  Our work
here uses general principles but focuses on the kink, for which exact  
results are not
available.  Dashen et al. put the object (classical background field  
corresponding to
kink or to s-G soliton) in a box of length
$L$ to discretize the
  continuous  spectrum, and used mode number regularization (equal  
numbers of modes in
the topological and trivial sectors, including the zero mode in
the topological sector in this counting) for the ultraviolet
divergences. They imposed periodic boundary conditions (PBC) on
the meson field which describes the fluctuations around the
trivial or topological vacuum solutions, and added a
logarithmically divergent mass counterterm whose finite part
  was fixed by requiring absence of tadpoles in the trivial background.  
They found for
the kink
\begin{equation} M^{(1)} =
\sum \frac{1}{2} \hbar \omega_n - \sum \frac{1}{2}
\hbar
\omega_n^{(0)} +
\Delta M = - \hbar m \left( \frac{3}{2 \pi} -
\frac{\sqrt{3}}{12}
\right) <0
\label{mass}  \ \ ,
\end{equation} where $m$ is the renormalized mass of the meson in the  
trivial background. This
result remains unchallenged.

The supersymmetric (susy) case, as well as the general case including  
fermions, proved
more difficult. The action and the kink solution read
$$ {\cal L}=- \frac{1}{2} \left(
\partial_{\mu}\phi
\right)^{2}-  \frac{1}{2}
\overline{\psi}\!\!\!\not\!\partial\psi -
\frac{1}{2} U^{2} - c\frac{1}{2}
\frac{dU}{d\phi}\overline{\psi}\psi
\ \ ,$$
\begin{equation}
\phi_{\rm kink}(x)=\frac{\mu}{\sqrt\lambda}\tanh\frac{\mu x}{\sqrt2} \  
\ ,
\end{equation} where $- \frac{1}{2} U^{2}=
-\frac{\lambda}{4}(\phi^{2}-\mu^{2}/\lambda)^{2}
$,  the meson mass is
$m=\mu\sqrt{2}$, and $c=1$ for supersymmetry. Dashen et al. did not  
explicitly compute
the fermionic corrections to the soliton mass, stating ``The actual  
computation of [the
contribution to]
$M^{(1)}$ [due to fermions] can be carried out along the lines of the  
Appendix. As the
result is rather complicated and not particularly illuminating we will  
not give it
here'' (page 4137 of
\cite{dashen}).

Several authors later computed $M^{(1)}$ for the susy kink and found  
different
answers. It became clear that the method of Dashen et al. yielded  
results that
depended on the boundary conditions  for the fluctuating fields. In  
fact, repeating
exactly the same steps for the susy kink as taken by Dashen et al. for  
the bosonic
kink (using PBC also for the fermions,
$\psi_{\pm}(-L/2) =\psi_{\pm}(L/2) $), taking equal numbers of modes in  
all four
sectors, including one term with
$\omega \simeq  0
$ (due to a periodic solution with $\omega^2>0$) in the bosonic kink  
sector and one
term with
$\omega=0$ in the fermionic kink sector (explicitly, there are two real  
independent
solutions with
$\omega=0$, one localized at the kink and  one at the boundary, and the  
coefficient of
each satisfies
$c^2=\frac{1}{2}$ \cite{fred})
  gives
  $M^{(1)}=\hbar m (\frac{1}{4} -
\frac{1}{2 \pi})$ \cite{rebhan}.  We now know that this $M^{(1)}$
is the correct answer to an inappropriate question, because it
includes boundary energy. Schonfeld \cite{schonfeld} finessed the
problem of a single kink with its sensitivity to boundary
conditions by considering the kink-antikink system with PBC.
Taking into account two terms with $\omega \sim 0$ in the bosonic
kink-antikink sector (due to one periodic solution with
$\omega^2<0$ and another antiperiodic solution with $\omega^2>0$)
and one term with $\omega= 0$ in the fermionic kink-antikink
sector (due to two periodic solutions with $\omega=0$
\cite{fred}),  he obtained what we now know to be the correct
answer. The problem of boundary contributions was circumvented by
other methods in \cite{misha} and \cite{graham}. The fermionic
contribution to $M^{(1)}$ is given by \begin{equation}
M^{(1)}_f=\hbar m\left(\frac{1}{\pi}-\frac{\sqrt{3}}{12}\right)>0
\ \ ,
\end{equation} and the total one-loop correction is thus
\begin{equation} M^{(1)}_{susy}=-\frac{\hbar m}{2\pi}
\label{Ms}\ \  .
\end{equation}

  With attention restricted to the kink alone, it was shown in  
\cite{fred} that  one
could eliminate boundary contributions from the fermionic part of the  
energy by
averaging over quartets of boundary  conditions for the fermionic  
fluctuations --
periodic, antiperiodic, twisted periodic and twisted antiperiodic,  
where twisting
means interchange of the upper and lower components of the fermion wave  
function
$\psi_{\pm}\to\psi_{\mp}$ \cite{misha}.  This averaging
  is necessary to preserve certain discrete symmetries for fermions. The
 results in
\cite{fred} give a complete, though  intricate,  way to
calculate
$M^{(1)}$ in terms of mode frequencies $\omega_n$.

In light of the complexities which boundary conditions generate
for the problem including fermions, the most important advance
since \cite{dashen}  was the approach of Shifman, Vainshtein, and
Voloshin \cite{shifman},  who used higher  space-derivative
regularization (with factors $(1-\partial_x^2/M^2)$ for the
kinetic terms but not the  interactions) to compute the central
charge densities of the susy sine-Gordon soliton and kink.  Their
scheme is manifestly susy, canonical (no  higher time
derivatives), and independent of  boundary conditions  (because it
yields a local density).\footnote{ As \cite{shifman} point out,
Yamagishi \cite{yam} was perhaps the first to study local
densities in this context.}  They argued that the energy density
is equal to the central charge density (because the difference is
a susy commutator) and they computed  the latter  -- including an
anomaly recognizable as an $M^2/M^2$  effect.  They  verified that
the one-loop correction $Z^{(1)}$ to the integrated central charge
of the kink comes only from the anomaly and is equal to
(\ref{Ms}). The presence of a topological anomaly was first
conjectured in \cite{misha}.

  One may compute the energy density for the bosonic
sine-Gordon soliton by mapping the system onto another one which  
exhibits supersymmetry,
and computing the central charge density
for that fictitious system
\cite{shifman}.  Quite possibly similar  techniques would work for the  
bosonic
kink.  Our approach here is instead to attack the
  Casimir energy density directly, freeing the calculation from  
dependence on
supersymmetry.  In doing so, we have found it helpful to introduce
a simple rule for regularization of energy densities which we
propose as a fundamental principle.  At this point, the primary
evidence for the validity of the principle is the agreement
between the energy density we compute by its use with the central
charge density of \cite{shifman}.  Our proposed
  principle appears new in the literature, and yet has roots in
early quantum physics. {\it Local mode regularization} (lmr), or
mode density regularization, is the local counterpart of the
familiar (global) mode regularization or mode number
regularization.   The principle is that, when fluctuations are
expanded into normalized modes $\phi_n(x)$ (i.e., $\int dx
\phi^*_n(x)\phi_n(x) =1 \  {\rm for \ all} \ n$), the cut-off
local mode density $\rho_N(x) = \sum_{n=1}^N \phi^*_n(x)\phi_n(x)
$ should be background-independent, meaning the same in the
trivial ($\rho^{(0)}$) and kink ($\rho$) sectors. For the bosonic
kink this implies that one must truncate the sums at different
upper bounds corresponding to different frequency cutoffs
\begin{equation}
\rho_\Lambda(x) = \rho_{\Lambda+\Delta \Lambda(x)}^{(0)} (x) \ \ .
\label{eq:eq3}
\end{equation}
Here the $x$-independent cutoff $\Lambda$ is given by
$\Lambda=2\pi N/L$
to an accuracy ${\cal O}(1/L)$.
The above equation determines $\Delta \Lambda(x)$ in terms of
$\Lambda$, and clearly
$\Delta \Lambda(x) $ is
$x$-dependent. For the susy kink one can begin by fixing the mode  
number  cut-off
$N_b^{(0)}$ for the bosons and
$N_f^{(0)}$  for the fermions in the trivial sector such that here the  
bosonic and
fermionic mode densities
  are equal. From (\ref{eq:eq3}) one then obtains in  the kink sector the
requirement
\begin{equation}
\rho_{\Lambda,b}(x) = \rho_{\Lambda+\Delta \Lambda(x),f} (x) \ \ ,  
\label{six}
\end{equation} which again determines $\Delta \Lambda(x) $ in terms of
$\Lambda$.  We use this principle to compute the anomalous energy
density of the bosonic kink, as well as of the supersymmetric
kink, which as mentioned was obtained already in \cite{shifman}
through the equality of energy and central charge densities.  We
believe that lmr is sufficient for regularization of the one-loop
Casimir energy density in one space dimension, and at least
necessary in higher dimensions, where further requirements may be
needed to specify the regularization completely. It is important
to emphasize that lmr, like ordinary mode regularization, is not
easily applied at arbitrary order in perturbation theory, because
it is not directly expressed in terms of a modification of the
action.  Thus our claim is that for a specialized purpose, namely,
calculation of localized Casimir energy, lmr is the ideal tool,
providing maximal simplicity and efficiency.  This claim is
simultaneously modest, because it is restricted to the computation
of one-loop energy densities, and substantial, because Casimir
energy plays such a large role in quantum physics.

For the non-anomalous contributions to both the bosonic and susy kink  
densities, we
find empirically another striking regularity, {\it phase space  
factorization}.  The
continuum contribution to the Casimir energy density in phase space   
exhibits a
remarkable  factorization, involving a few terms each with simple  
momentum-dependent
factors multiplying functions related to the bound-state and zero-mode  
probability
densities in coordinate space.  We believe
this factorization should hold for all reflectionless potentials, but  
might not extend farther.
  The factorization takes a particularly simple form for
what we call below $\epsilon_{\rm Cas}(x)$, a local density whose
integral over a region containing the kink gives the total quantum
correction to the mass of the kink. The local energy density has
two contributions besides $\epsilon_{\rm Cas}$. First, the
fundamental definition of Casimir energy density differs from the
usual sum over zero-point energies by an extra piece which is a
perfect differential of an expression vanishing far from the kink,
so that including this piece does not alter the mass correction
but does alter the local density.  Secondly, there is an effect
which perhaps is best viewed as vacuum polarization -- the quantum
fluctuations of the Bose and Fermi fields lead to a local shift in
the classical background field defining the kink.  Again, this
changes the local density but not the total mass. We check
explicitly that the energy density and the central charge density
of the susy kink are equal.

After these calculations of the anomalous and non-anomalous
contributions to the energy density of the bosonic (and susy)
kink, we turn to the calculation of the central charge density of
the susy kink.  From its definition the central charge density at
a point $y$ is an $x$-integral of a bilocal quantity depending on
$x$ and $y$ times a delta function $\delta(x-y)$.  Not setting
$x=y$ too soon yields the anomaly in the central charge density
  near the kink, confirming \cite{shifman}. Finally, we discuss the
physical basis for lmr, observing in particular that point-splitting  
regularization
for energy density implies lmr, at least for the bosonic case with arbitrary
background potential.

{\it  Bosonic kink energy density.---}For the energy density of the  
bosonic kink, one
must evaluate sums (setting $\hbar=1$ from now on)
$\sum \frac{1}{2}
\omega_n
\phi^*_n(x) \phi_n(x) $, where the modes $\phi_n(x) $ are each  
normalized to unity.  As these
sums clearly differ from the density sums $\sum
\phi_n^*(x)
\phi_n(x)$,  one expects in general a nonvanishing one-loop correction  
to the energy
density, and hence to the quantum mass.  Let us begin with explicit  
expressions for
the mode eigenfunctions,  so that one may follow the argument in  
detail.  The wave
functions of the continuous spectrum (using  $ |\phi_n|^2(x)=1$ away  
from the kink to
determine the normalization constant ${\cal N}$) obey
\begin{equation}
\phi(k,x) = \frac{e^{ikx}}{\cal N} \left[ -3 \tanh^2
\frac{mx}{2}+1+ 4
\left(
\frac{k}{m} \right)^2 + 6i \frac{k}{m} \tanh
\frac{mx}{2} \right] \ \ ,  \label{m7}
\end{equation} with $\omega= \sqrt{k^2+m^2} $  and
$ {\cal N}^2= 16 \frac{\omega^2}{m^2} \left(
\frac{\omega^2}{m^2} - \frac{\omega_B^2}{m^2}
\right)$. The zero mode with
$\omega_0=0$ is given by
\begin{equation}
\phi_0(x) = \sqrt{\frac{3m}{8}} \frac{1}{\cosh^2 (mx/2)} \ \ .
\end{equation} The bound state with $\omega_B =
\frac{\sqrt{3}}{2} m$ is given by
\begin{equation}
\phi_B(x) =\sqrt{\frac{3m}{4}} \frac{\sinh  (mx/2)}{\cosh^2 (mx/2)} \ \
. \label{m9}
\end{equation} The density of the continuous spectrum  can be written  
as follows
$$ |\phi(k,x)|^2 = \frac{1}{{\cal N}^2} \left[
\frac{4 \omega^2}{m^2} \left( \frac{4
\omega^2}{m^2}-3
\right) - 12 \  \frac{\omega^2}{m^2} \frac{1}{\cosh^2
  (mx/2)} + 9 \ \frac{1}{\cosh^{4}
  (mx/2)}
\right]
$$
\begin{equation}  = 1 - \frac{m
\phi_B^2(x)}{\omega^2-\omega_B^2}-
\frac{2 m \phi_0^2(x)}{\omega^2} \ \
\label{eq10m}
\end{equation}
  As $ \int_{-\infty}^{\infty} \frac{dk}{2 \pi }
\frac{1}{\omega^2-\omega_B^2} = \frac{1}{m} $, while
$
\int_{-\infty}^{\infty} \frac{dk}{2 \pi }
\frac{1}{\omega^2} =
\frac{1}{2m} $, it is clear that the completeness relation
\begin{equation}  \ \ \ \ \
\int_{-\infty}^{\infty} \frac{dk}{2 \pi} \left\{ |\phi(k,x)|^2-1
\right\}+ \phi_0^2(x) + \phi_B^2(x) =0
\label{comp}
\end{equation}  is satisfied.  Eq. (\ref{eq10m}) may be written in a  
remarkable
formula perhaps true for all reflectionless potentials,  showing  
factorization of the
difference in mode densities in phase space, where the position  
dependence of each
term is given by the corresponding bound-state or zero-mode probability  
density,

\noindent
  {\bf Relation 1}:
\begin{equation} |\phi(k,x)|^2 - 1 = - \sum_j \phi_{j}^2(x)
\frac{2
\sqrt{m^2-\omega_{j}^2}}{\omega^2-\omega^2_{j}} \ \ ,
\end{equation} satisfying the completeness relation, as one may check  
by performing
the integration over
$k$.

Note that all the above expressions for the density do not refer
to any particular choice of boundary conditions, which of course
do affect eigenenergies and the corresponding wave  functions. The
reason  is that the choice of boundary conditions will contribute
to the density away from the boundary at most terms of order
$1/L$. In the large-$L$ or continuum limit, in principle such
terms might contribute to the total energy obtained by integration
over the entire interval between the boundaries. We are unaware of
any example of this phenomenon, as a previous claim in \cite{fred}
has been shown to be incorrect \cite{grvw}.\footnote{However, the
latter work did identify a delocalized {\it momentum} for certain
special boundary conditions.}
  Even if the phenomenon were to occur, for the integral
just over a finite interval around the kink the effect would be
negligible, so that the kink energy density and resulting energy
can be computed reliably in terms of the continuum,
modified-plane-wave solutions, unaffected by the choice of
boundary conditions.

The requirement in (\ref{eq:eq3}) that the topological vacuum density  
and the trivial
vacuum density be equal leads via (\ref{comp}) to
$$
\frac{\Delta \Lambda(x)}{\pi} =
\int_\Lambda^\infty \frac{dk}{\pi} \left( 1-|\phi(k,x)|^2
\right)
  =\int_\Lambda^\infty dk\frac{2}{\pi}\sum_j\frac{\sqrt{m^2-
\omega^2_{j}}}{\omega^2-
\omega^2_{j}}
\phi_{j}^2(x)$$
\begin{equation} = \frac{m}{\pi \Lambda} \left( \phi_B^2(x) + 2
\phi_0^2(x)
\right) +{\cal O} \left( \frac{1}{\Lambda^2}
\right)= \frac{3 m^2}{4 \pi \Lambda} \frac{1}{\cosh^2
  (mx/2) } + {\cal O} \left( \frac{1}{\Lambda^2}
\right)\label{deltalambda} \ \ .
\end{equation}

With this result for $\Delta \Lambda(x)$ we can evaluate the energy  
density
$\epsilon(x)$ in the kink sector. Adding also the counterterm\footnote{  
The counterterm $\Delta
M(x)$ usually is expressed in terms of the kink background field, but  
it also can be
determined by noting that it should cancel the remaining divergence in  
the
  difference of the sums over zero point
energies $\int_{-\infty}^\infty
(dk/2\pi)|\phi(k,x)|^2(-\frac\omega 2)$.  Equating both
expressions yields another formula perhaps valid for general
reflectionless potentials,

\noindent {\bf  Relation 2}:
  $$ \frac{\sum_j 2\phi_j^2(x)\sqrt{m^2-\omega_j^2}}{\sum_j  
2\sqrt{m^2-\omega_j^2}} =
\frac{\phi_K^2(x) -\phi_K^2(\infty)}{\int_{-\infty}^\infty dx[
\phi_K^2(x) -\phi_K^2(\infty)]}\ \ .$$}

  \begin{equation}\Delta M(x) =
\sum_j 2\phi_{j}^2(x)\sqrt{m^2-\omega_{j}^2}
\int_0^{\Lambda}\frac{dk}{2\pi}\frac{1}{\omega}=
\frac{m^2}{4} \frac{1}
{\cosh^{2}  (mx/2)}\frac{3}{2 \pi}
\int_0^\Lambda
\frac{dk }{\omega} \ ,
\end{equation} and rewriting $\frac{1}{\cosh^{2}
  (mx/2)}$ as $
\frac{4}{3m} \phi_B^2(x) + \frac{8}{3m} \phi_0^2(x)$ yields
$$\epsilon_{\rm Cas}(x)=
\epsilon(x) - \epsilon^{(0)} (x) = \frac{1}{2}
\omega_B
\phi_B^2(x) +  2\int_0^\Lambda \frac{dk}{2 \pi} |\phi(k,x)|^2
\frac{1}{2} \omega - 2 \int_0^{\Lambda+\Delta
\Lambda(x)}
\frac{dk}{2\pi}  \frac{1}{2} \omega + \Delta M(x)
  =$$
\begin{equation} \frac{1}{2} \omega_B \phi_B^2(x) -
\int_0^\Lambda
\frac{dk}{2 \pi}
\left(
\frac{m \phi_B^2(x)}{k^2+ m^2/4} + \frac{2 m\phi_0^2(x)}{k^2+m^2}
\right)
  \omega
  - \frac{ \Delta \Lambda (x)}{2\pi}
\Lambda + m \left(
\phi_B^2(x) + 2 \phi_0^2(x) \right) \int_0^\Lambda
\frac{dk}{2 \pi}
\frac{1}{\omega} \ \ .
\end{equation} The two quadratic divergences proportional to
$\int_0^\Lambda dk \ \omega$ have canceled because we subtracted the  
energy density of
the trivial vacuum, while the counter term cancels the remaining  
logarithmic
divergence.  Again, each term is proportional to a bound-state or  
zero-mode probability density.

The result is finite and reads
\begin{equation}
\epsilon_{\rm Cas}(x) =
\frac{1}{2}
\omega_B \phi_B^2(x) - m
\int_0^\Lambda
\frac{dk}{2 \pi}
\left(
\frac{\omega}{k^2+ m^2/4} -
\frac{1}{\omega} \right) \phi_B^2(x) -
\frac{m}{2 \pi} \left(\phi_B^2(x)+2 \phi_0^2(x)
\right) \ \ .\label{bonomaly}
\end{equation} The last term is the contribution from the term due to
$\Delta \Lambda (x) $, and is the analogue of the term in the central  
charge density
for the susy case identified as the anomaly by \cite{shifman}.
  Using the integral
\begin{equation}
\int_0^\infty \frac{dk}{2 \pi} \left(
\frac{1}{k^2+ m^2/4} -
\frac{1}{k^2+m^2}
\right) \omega = \frac{1}{2 \sqrt{3} } \ \ ,\label{intcas}
\end{equation} we obtain
\begin{equation}
\epsilon_{\rm Cas}(x) = \left(
\frac{1}{2} \omega_B - \frac{m}{2 \sqrt{3}} -
\frac{m}{2 \pi}
\right)
\phi_B^2(x) - \frac{m}{\pi} \phi_0^2(x) \ \ . \label{epscas}
\end{equation}  This formula can be rewritten as  follows,

\bigskip

\vspace{-.15in}
\begin{equation}
\epsilon_{\rm Cas}(x)=
\sum_j \frac{1}{2} \left( 1 - \frac{2}{\pi} \arctan
\frac{\omega_j}{\sqrt{m^2- \omega_j^2}} \right)
\omega_j\phi^2_{j}(x)-
\sum_j\frac{1}{\pi}\sqrt{m^2-
\omega^2_{j}}
\phi_{j}^2(x) \ \ , \label{fad} \end{equation} where in the first sum  
the contribution
with 1 comes from the bound states, and that with arctan comes from the  
continuum,
while the second sum is the anomaly contribution.  Such formulas for  
the total mass
can be found in
\cite{Faddeev}, though we are unaware of local versions in the  
literature.  This kink
example might be an illustration of a general factorization rule, valid  
for a wide
class of reflectionless potentials.  While we have not tested it for  
other cases, and
do not know how to prove it other than by explicit computation, we  
believe that its
simplicity and elegance make the rule worthy of further investigation.

Integration of $\epsilon_{\rm Cas}(x)$ over $x$ yields
\begin{equation} M^{(1)}= \frac{1}{2}
\omega_B(1-\frac{2}{3})-
\frac{m}{2\pi} -\frac{m}{\pi} =
\frac{\sqrt{3} m}{12} -
\frac{3m}{2
\pi} \ \ ,
\end{equation} in agreement with (\ref{mass}).
For convenience later, let us express (\ref{epscas}) as a total  
derivative:
\begin{equation}
\epsilon_{\rm Cas}=\frac
m2\frac{d}{dx}\bigg{\{}\tanh(mx/2)\bigg{[}\frac{\sqrt3}{12}\tanh^2(mx/ 
2)-\frac{3}{2\pi}\bigg{]}\bigg{\}}
\ \ . \label{casint}\end{equation} The first term gives the
non-anomalous contribution, while the second yields that due to
the anomaly.

Eqs. (\ref{fad}, \ref{casint}) give compact expressions for the
local energy density, which certainly provide the correct total
quantum energy of the bosonic kink.  However, to obtain the
correct local Casimir energy density, one must start with an
expression for the energy density of each mode including a
quadratic term in the gradient of the boson field
$\frac{1}{2}(\partial_x\eta)^2$, whereas our formulae above,
giving energies $\omega_n$ multiplying corresponding mode
probability densities, implies instead the expression coming from
the field equation,
$-\frac{1}{2}\eta\partial_x^2\eta$.\footnote{Our boson fluctuation
field $\eta$ is designated by $\chi$ in \cite{shifman}.  Note
that, as utilized just below, for a mode of frequency $\omega_n$
our normalized mode function is given by
$\phi_n=\sqrt{2\omega_n}\eta_n$.} Therefore we need to add to
(\ref{fad}) the difference, a perfect differential of a function
which vanishes far from the kink (and so does not change the
computed mass correction  for the kink),
\begin{equation}
\Delta\epsilon_{\rm Cas} (x)=
\frac{1}{4}\partial_x^2\langle\eta^2(x)\rangle
\ \ .  \label{twn}
\end{equation} The propagator at equal times and positions
$\langle\eta^2(x)\rangle$ (excluding the zero mode which solves the  
homogeneous
equation) \cite{shifman} can be obtained by integrating  
$\frac{|\phi|^2}{2\omega}$ in
(\ref{eq10m}) w.r.t. $\frac{dk}{2\pi}$ and adding  
$\frac{\phi_B^2}{2\omega_B}$. The
divergent part in $\langle\eta^2\rangle$ corresponds to setting $|\phi  
(k,x)|^2$ equal
to its value at $|x|\to \infty$ in the $k$ integral; as the divergence  
is
$x$-independent it cancels in (\ref{twn}).
\smallskip
\noindent   One thus obtains a finite (because the logarithmically
divergent part has been subtracted) expectation value
$$\langle\eta^2(x)\rangle_{\rm  
r}\equiv\langle\eta^2(x)-\eta^2(x=\infty)\rangle =
\int_{-\infty}^\infty
\frac{dk}{2\pi}\frac{1}{2\omega}[|\phi(k,x)|^2- 
1]+\frac{1}{2\omega_B}\phi_B^2(x)=$$
\begin{equation} =-\frac{3}{8\pi}\frac{1}{\cosh^4(mx/2)}+
\frac{1}{4\sqrt{3}}\frac{\sinh^2(mx/2)}{\cosh^4(mx/2)} \ \  
\label{twentyone},
\end{equation}
and therefore,
\begin{equation}
\Delta\epsilon_{\rm Cas}=\frac{d}{dx}\bigg{\{}\tanh
(mx/ 
2)\bigg{[}\frac{m}{4}\bigg{(}\frac{3}{4\pi}+\frac{1}{2\sqrt3}\bigg{)}
\frac{1}{\cosh^4(mx/2)}
-\frac{m}{16\sqrt3}\frac{1}{\cosh^2(mx/2)}\bigg{]}\bigg{\}} \ \ .
\end{equation}

As observed in \cite{shifman}, besides the Casimir energy density there  
is another
consequence of the zero-point oscillations, namely, a  
position-dependent shift
$\phi_1$ in the classical  background field.  This in turn implies a  
further term in
the local energy density, given by
\begin{equation}
\Delta \epsilon_{(
\phi_1)}(x)=\partial_x\phi_1\partial_x\phi_{\rm  
kink}+(\frac{1}{2}U^2)'\phi_1=
\partial_x(\phi_1\partial_x\phi_{\rm kink})  \ \ ,\label{ephi}
\end{equation} but of course no shift at this order in the total  
energy, because the
classical energy is stationary with respect to arbitrary small           
  variations of
the  classical field about its equilibrium form.  Decomposing the  
Heisenberg field
$\Phi(x,t)$ as $\phi_{\rm kink}(x)+\phi_1(x)+\eta(x,t)$, with the  
quantum fluctuation
field obeying $\langle\eta\rangle=0$, and taking the expectation value  
of the $\Phi$
field  equation $\langle -\partial_t^2\Phi
+\partial_x^2\Phi-(\frac{1}{2}U^2)'\rangle=0$  gives
\begin{equation}
\partial_x^2\phi_1-(\frac{1}{2}U^2)''\phi_1=
\frac{1}{2!}(\frac{1}{2}U^2)'''\langle
\eta^2\rangle-\frac{1}{2}\Delta m^2\phi_{\rm kink}
\
\ .\label{toot}
\end{equation}
\noindent This $\phi_1$ is just what is needed to satisfy the  
no-tadpole condition
in the kink background. As mentioned above, the
singularity in
$\langle\eta^2(x)\rangle$ is $x$-independent and compensated by the
  $\Delta m^2$ term, yielding the quantity $\langle\eta^2(x)\rangle_{\rm
r}$ of
(\ref{twentyone}).\footnote{This procedure gives $\phi_1$ as a finite,  
renormalized quantity,
while
\cite{shifman} use an unrenormalized $\phi_1$. Hence the `rescaling'  
part of their $\phi_1$ gives
the shift to the pole mass from the unrenormalized mass, but ours gives  
only the shift to the pole
mass from the renormalized mass corresponding to the vanishing-tadpole   
condition in the
trivial sector.} Solving
(\ref{toot}) by the  Ansatz\footnote{The term with $A$ is  needed for  
the term in
(\ref{twentyone}) proportional to
$1/\cosh^2(mx/2)$ and the term with $B$ is needed for terms with    
$1/\cosh^4(mx/2)$.}
  $\phi_1=Ax\phi_0(x)+ B\partial_x\phi_0(x)$, and using the fact that  
$\phi_0$
is proportional to $\partial_x\phi_{\rm kink}$, one finds
\begin{equation}
\phi_1=\frac {\lambda}{m^2} {\bigg [ } \left
(\frac{1}{2\sqrt{3}}+\frac{3}{4\pi}\right )
\frac{1}{\cosh^2(mx/2)}
-\frac{\sqrt{3}}{4}(m\partial_m+2\lambda\partial_{\lambda}){\bigg ]  
}\phi_{\rm kink}
\
\ \label{phi1}.
\end{equation} Through one-loop order  (as in the susy case
\cite{shifman}), the effect of the second term is to replace
  the renormalized mass
$m$ and coupling $\lambda$ in $\phi_{\rm kink}$  with the bosonic pole  
mass $\bar
m=m(1-\sqrt{3}\frac{\lambda}{4m^2})$ given in
\cite{rebhan}, eq.(7) and the adjusted coupling $\bar \lambda =
\frac{\bar m^2} {m^2}\lambda$.  If we then rewrite the classical energy  
in terms of
$\bar m$ and $\bar \lambda$, the classical mass is multiplied by a  
factor
$1-\sqrt{3}\frac{\lambda}{4m^2}$. As $\phi_1$ cannot shift the total  
mass, we know
even without explicit calculation that the classical energy density in  
terms of the
barred quantities must be renormalized by a compensating factor
$1+\sqrt{3}\frac{\lambda}{4m^2}$.  The first term in (\ref{phi1}) is  
sensitive only to
bosonic fluctuations and hence unchanged in the susy case (because, as  
we shall see,
the fermionic source for  $\phi_1$ includes no terms with
$1/\cosh^4 (mx/2)$); it contributes according to
(\ref{ephi}). The total one-loop bosonic energy density becomes

\noindent {\bf Relation  3}:
$${\cal E}(x) = U^2(\bar\lambda,\bar m,\phi_{\rm kink}(\bar \lambda,
\bar m,x)){\bigg ( }1+\sqrt{3}\frac{\bar\lambda}{4\bar m^2}{\bigg ) } \  
+
$$
\begin{equation}+ \ \epsilon_{\rm Cas}(x)+\Delta \epsilon_{\rm Cas}(x)  
+ \frac{\bar
m}{4}{\bigg ( }
\frac{3}{4\pi} +\frac{1}{2\sqrt{3}} {\bigg ) }\partial_x {\bigg (  
}\frac{{\rm tanh}
(\bar m x/2)}{\cosh^4 (\bar m x/2)} {\bigg ) } \ \   \label{25m} .
\end{equation} The last term in (\ref{25m}) comes from the first term  
in (\ref{phi1}).
The second term in (\ref{phi1}) renormalizes the classical contribution  
$U^2$, as seen
in the first term in (\ref{25m}). The effect of this renormalization is  
that the classical
energy density flattens out a bit.  Besides the rescaling, all other  
contributions are of
the form $\partial_x[\phi_{\rm kink}(x)/\cosh^n(mx/2)]$, with $n=$  
0,2,4.

{\it Susy kink energy density.---} For the susy kink we choose the
cut-offs in the trivial sector in such a way that the bosonic and
fermionic densities in that sector are equal. To make the bosonic
and fermionic densities also equal in the topological sector, we
use a cut-off $\Lambda$ for the bosons and $\Lambda+\Delta
\Lambda(x)$ for the fermions as in (\ref{six}). The fermion is
described by a Majorana two-component spinor $\psi={\psi_+ \choose
\psi_-}$. As $\psi_+(k,x)$ is proportional to $\phi(k,x)$ while
$\psi_-(k,x) = \frac{i}{\omega} \left( \partial_x + m \tanh
\frac{mx}{2} \right) \psi_+(k,x) $  for solutions proportional to
$\exp (-i \omega t)$ according to the Dirac equation
\cite{rebhan}, one obtains for the wave functions of the
continuous fermionic spectrum
\begin{equation}
\psi_+(k,x) = \frac{1}{\sqrt{2}} \phi(k,x)
\end{equation}
\begin{equation}
\psi_-(k,x) = \frac{1}{\sqrt{2}} \frac{\omega}{{\cal N} m}
\left( - 4
\frac{k}{m} - 2 i \tanh \frac{mx}{2} \right) e^{ikx}
\ \ .\end{equation} In the difference of the densities the constant  
term of course
cancels, giving
$$ |\phi(k,x)|^2 - |\psi_+(k,x)|^2 - |\psi_-(k,x)|^2 = |\psi_+(k,x)|^2-
|\psi_-(k,x)|^2 \
\
$$
\begin{equation} = \frac{1}{2 {\cal N}^2} \left(\frac{ 9}{\cosh^{4}
  (mx/2)} - 8 \left(
\frac{\omega}{m} \right)^2 \frac{1}{\cosh^{2}
  (mx/2)} \right) \ \ .  \label{asym}
\end{equation} For the bosonic and fermionic densities to satisfy  
(\ref{six}) one
requires
$$
\phi_0^2(x) + \phi_B^2(x) + 2 \int_0^\Lambda
\frac{dk}{2 \pi}  |\phi(k,x)|^2
$$
\begin{equation} = \frac{1}{2} \phi_0^2(x)+ \left (\frac{1}{2}
\phi_B^2(x) +
\frac{m}{8 \cosh^2  (mx/2) }\right ) + 2
\int_0^{\Lambda +
\Delta \Lambda} \frac{dk}{2\pi}\left\{ |\psi_+(k,x)|^2+ |\psi_-(k,x)|^2
\right\} \ \ .
\end{equation}  The factor $\frac{1}{2}$ in
$\frac{1}{2}\phi_0^2(x)$ comes from the mode expansion
$\psi_+(x,t)=c_0\phi_0(x,t)+{\ ...}$ with $\{ c_0,c_0\}=1$.
\footnote{Obtained in \cite{fred} from Dirac
quantization, this factor equivalently can be deduced from the  
completeness
relation for solutions of the single-particle Dirac equation
$\phi_0^2(x)+2[|\psi_{B+}|^2+|\psi_{B- 
}|^2+\int\frac{dk}{2\pi}(|\psi_+|^ 2
+|\psi_-|^2-1)]=0$, where
$|\psi_-(k,x)|^2=1/2-m^2/(8\cosh^2(mx/2)(\omega^2-\omega_B^2))$.
The relative factor of two between the first term and the later terms  
in the completeness
relation follows from the fact that, if one sums over a complete set of  
solutions of the Dirac
equation, all nonzero frequencies lead to equal contributions from  
positive and from negative
frequency, while the zero mode contributes only once. } This
$\frac{1}{2}$ is the analogue for Majorana fermions of the fractional  
fermion charge discovered
by Jackiw and Rebbi for Dirac fermions
\cite{JR}.
  The two terms in parentheses give the
$\psi_+$ and
$\psi_-$ contributions of the bound state:
$\psi_{B+}^2=\frac{1}{2}\phi_B^2$ and
$|\psi_{B-}|^2=\frac{m}{8\cosh^2(mx/2)}$.
   We obtain
$$
\frac{1}{2} \phi_0^2(x) +  \psi_{B+}^2(x)  -|\psi_{B-}|^2(x)+ 2
\int_0^\Lambda
  \frac{dk}{2 \pi}  \left( |\psi_+(\Lambda ,x)|^2 - |\psi_-(\Lambda ,  
x)|^2 \right)
$$
\begin{equation}      =\frac{\Delta \Lambda(x)}{\pi} \left(  
|\psi_+(k,x)|^2 +
|\psi_-(k,x)|^2 \right) \ \ .
\end{equation} Using the completeness relation, and taking the large  
$k$ limit $
|\psi_+(k,x)|^2 + |\psi_-(k,x)|^2
  \to 1$, one finds
\begin{equation}
\frac{\Delta \Lambda(x)}{\pi} = -2
\int_\Lambda^\infty
  \frac{dk}{2\pi}  \left(|\psi_+(k,x)|^2 - |\psi_-(k,x)|^2
\right) \ \ .
\end{equation} As we are interested only in the $1/\Lambda$ term, the  
calculation is
easy.  From (\ref{asym}) we find
\begin{equation}
\frac{\Delta \Lambda (x)}{\pi} = \frac{m^2}{4\pi\Lambda}
\frac{1}{\cosh^2(mx/2)} \ \ .
\end{equation}

With this result in hand, we compute the difference in energy densities  
for the susy
kink
$$
\epsilon_{\rm Cas,b}(x) - \epsilon_{\rm Cas,f}(x) =
\frac{1}{2}
\omega_B \left(\phi_B^2(x)-\psi_B^{\dagger}(x)
\psi_B(x)
\right)
$$
\begin{equation} + 2 \int_0^\Lambda \frac{dk}{2 \pi}  \left(  
|\psi_+(k,x)|^2 -
|\psi_-(k,x)|^2 \right) \frac{1}{2}
\omega -
\frac{\Delta \Lambda(x)}{\pi} \frac{1}{2} \Lambda +
\Delta M_{\rm susy}(x)  \ \ .
\end{equation} The counter term in the susy case,
\begin{equation}
\Delta M_{\rm susy}(x) = \frac{m^2}{2} \frac{1}{\cosh^2
  (mx/2) }
\int_0^\Lambda
\frac{dk}{2 \pi} \frac{1}{\omega} \ \ ,
\end{equation}  is a factor 1/3 smaller, but still nonvanishing.
   Again, the counter term removes the logarithmic divergence in the  
integral, and with
(\ref{intcas}) one finds
\begin{equation}
\epsilon_{\rm Cas,susy}(x)=\epsilon_{\rm b}(x) - \epsilon_{\rm
f}(x)= (1-\frac{2}{3})\frac{1}{2} \omega_B
\left(\phi_B^2(x)-\psi_B^{\dagger}(x) \psi_B(x)\right) -
\frac{m^2}{8 \pi } \frac{1}{\cosh^2 (mx/2)} \ \ , \label{casu}
\end{equation} where the last term, the contribution from  
$\Delta\Lambda$, agrees with the
central charge density anomaly eq.(3.38) in
\cite{shifman}. Integration over $x$ yields the one-loop correction to  
the mass of the
susy kink
$$ M^{(1)}_{\rm susy} = lim_{X\to\infty}\int_{-X}^X dx\frac{d}{dx}
\bigg{[}\frac{\sqrt3m}{48}((\tanh^3(mx/2)-\tanh(mx/2))- 
\frac{m}{4\pi}\tanh(mx/2)\bigg{]}$$
\begin{equation}=
  \frac{1}{6} \omega_B(1-1) -
\frac{m}{2 \pi} =-
\frac{m}{2 \pi} \ \ , \label{casual}
\end{equation}   which of course is the accepted answer. Note that the  
non-anomalous
contributions from the bosons and the fermions do not cancel locally,  
but in the
integral they do:
$\frac{1}{6}
\omega_B(1-1)=0$. For explicit expressions later it is helpful to  
rewrite the first
part in the bracket of (\ref{casual})
as \begin{equation} \epsilon_{\rm Cas,susy}({\rm non-anom}) =  
\frac{d}{dx}
\frac{m}{16\sqrt3}(-\tanh(mx/2)/\cosh^2(mx/2)) \ \ .
\end{equation}

  As in the bosonic case we must add the missing term in the bosonic
Casimir energy density
$\Delta \epsilon_{\rm Cas}$ given in eqs. (\ref{twn},\ref{twentyone}),  
as well as include the
shift for the susy case
$\phi_1(x)$ in the background field.  We compute this $\phi_1$, again  
using the
second-order field equation for $\Phi$, which now has an additional  
contribution from
fermions:
\begin{equation}
\partial_x^2\phi_1-(\frac{1}{2}U^2)''\phi_1=
\bigg{\{}\frac{1}{2!}(\frac{1}{2}U^2)'''\langle
\eta^2\rangle-\frac{1}{2}\Delta m_b^2\phi_{\rm kink}\bigg{\}}+
\bigg{\{}\frac12U''\langle\bar{\psi}\psi\rangle-\frac{1}{2}\Delta  
m_f^2\phi_{\rm kink}\bigg{\}}
\
\ ,\label{toot'}
\end{equation}
with
$$\langle \bar{\psi}\psi\rangle =  
\int^\infty_{-\infty}\frac{mdk}{2\pi\omega}
[-1+6m^2/\cosh^2(mx/2)]\tanh(mx/2)/16(k^2+m^2/4)]$$
\begin{equation}
  -\sqrt3m/4\tanh(mx/2)/\cosh^2(mx/2) \ \ .
\end{equation}
The last term comes from the bound state, and the term with  
$(k^2+m^2/4)$ in the numerator of
the integrals is cancelled by the fermionic part of the mass counter  
term.  Again using
(\ref{intcas}), we find that the fermionic contributions to $\phi_1$  
are only proportional
to $1/\cosh^{2}(mx/2)$, and not  $1/\cosh^{4}(mx/2)$, so that the net  
coefficient of the
  $1/\cosh^{2}(mx/2)$ term is a factor $2/3$ smaller than in the  
bosonic case.  The final
result for $\phi_1$ reads
\begin{equation}
\phi_{1\rm ,susy}=\frac {\lambda}{m^2} {\bigg [ } \left
(\frac{1}{2\sqrt{3}}+\frac{3}{4\pi}\right )
\frac{1}{\cosh^2(mx/2)}
-\frac{1}{2\sqrt{3}}(m\partial_m+2\lambda\partial_{\lambda}){\bigg ]  
}\phi_{\rm kink}
\
\ \label{phi1s}.
\end{equation}
Note that in (\ref{phi1s}) the first term is the same as in the bosonic  
case (as mentioned
earlier), while the second term is smaller by a factor 2/3.
This is the same result found in \cite{shifman} using a first order  
differential equation
based on susy considerations.
Iterating the susy relation
$\langle\partial_x\phi+U\rangle=0$, one confirms that the second-order  
and first-order
approaches are consistent with each other.\footnote
  {The details are as follows.  From $\langle\partial_x\phi+U\rangle=0$  
we have
$\partial_x\phi_1+U'\phi_1+(1/2!)U''\langle\eta^2\rangle+\sqrt{\lambda/ 
2}(
-\Delta\mu^2/\lambda)=0 \ \ .$  Differentiating with respect to $x$,  
using $\partial_x\phi_{\rm
kink}=-U$, and eliminating $\partial_x\phi_1$ yields an equation for  
$\partial_x^2\phi_1$. That
this is equivalent to (\ref{toot'}) follows from the identity  
$\langle\eta(\partial_x+U')\eta +
((\partial_x+U')\eta)  
\eta\rangle=i\langle\psi_+\psi_--\psi_-\psi_+\rangle$, which in turn is  
a
consequence of $\psi_-=(i/\omega)(\partial_x\tanh(mx/2))\psi_+$ and  
$\psi_+=\eta\sqrt\omega$.
}
   For the susy energy density we then find full agreement with the  
central charge density of
\cite{shifman}, after  restoring a missing factor of
$\frac{1}{2}$ in the first line of (5.21) in that work, kindly pointed  
out to us  by
the authors.

\noindent {\bf Relation}  4: $${\cal E}(x) =
U^2(\bar\lambda_s,\bar m_s,\phi_{\rm kink}(\bar \lambda_s, \bar
m_s,x)){\bigg ( }1+\frac{\bar\lambda_s}{2\sqrt{3}\bar m_s^2}{\bigg
) } \ +
$$
\begin{equation}+ \ \epsilon_{\rm Cas,susy}(x)+\Delta \epsilon_{\rm  
Cas}(x) + \frac{\bar
m_s}{4}{\bigg ( }
\frac{3}{4\pi} +\frac{1}{2\sqrt{3}} {\bigg ) }\partial_x {\bigg (  
}\frac{{\rm tanh}
(\bar m_s x/2)}{\cosh^4 (\bar m_s x/2)} {\bigg ) } \ \  .
\end{equation}
Here we use $\bar{m}_s=m(1-\lambda/2\sqrt3m^2)$ and
$\bar{\lambda}_s=\lambda(1-\lambda/\sqrt3m^2)$. Thus the rescaling
of the classical density involves a shift 2/3 as big as for the
bosonic case, while $\epsilon_{\rm Cas,susy}$ includes an anomaly
1/3 as big as in the bosonic case and a different finite-energy
contribution, as explained above.  Meanwhile, the last two terms,
from $\Delta \epsilon_{\rm Cas}$ and from the nonrescaling term in
$\phi_{1\rm r}$, are the same as in the bosonic case.\footnote{For
comparison with \cite{shifman} note that our coupling $\lambda$ is
equal to $2\lambda^2$ in their formulation.  Also, they give the
integral of the density from $-x$ to $x$, while we write the
pieces of the quantum correction to the density as local
derivatives, so that our expression for the function being
differentiated is half theirs for the integral.  The specific
terms in their equation (5.21) are related to ours as follows: The
first line in (5.21) is simply the integral of what we call $U^2$.
The first term in the second line is the anomaly. In the final
bracket, the first term receives equal contributions from
$\epsilon_{\rm Cas}$ and $\Delta\epsilon_{\rm Cas}$.  The
remaining piece receives equal contributions from
$\Delta\epsilon_{\rm Cas}$ and from the nonrescaling part of
$\phi_1$, the shift in the classical field.}

{\it Central charge density.---}We now compute the anomalous  
contribution to the
density
$\zeta (x)$. Before  regularization one has
${\cal H}(x)=\frac{1}{2}\dot{\phi}^2 +\frac{1}{2}\phi'^2+\frac{1}{2}U^2+
\frac{i}{2}\left (
\psi_+\psi'_+ +
\psi_-\psi'_-\right )-iU'\psi_+\psi_-$, and
$\zeta (x) = U
\partial_x \phi$  (note that
\cite{shifman} have the opposite sign convention for
$\zeta$).  Using the equal-time anticommutators of the fermionic fields
$\psi_+(x)$ and $\psi_-(x)$ and the definition
\begin{equation} j_{\pm} = (-( \  
\!\!\!\not\!\partial\phi+U)\gamma^0\psi_{\pm} \ \  ,
\end{equation} one obtains
\begin{equation}
\{Q_{\pm}, j_{\pm}(y)\} = 2 {\cal H} (y)\pm 2 \zeta (y);
\ \ \ \  j_{\pm} = \dot{\phi} \psi_{\pm} + (
\phi^\prime
\pm U) \psi_{\mp};
\ \ \ \  Q_\pm = \int_{-L/2}^{L/2} j_{\pm} dx \ \ ;
\end{equation}
$$
\zeta(y) = \int dx \left[ \frac{1}{2} \left(
\{ \psi_+(x), \psi_+(y) \} - \{ \psi_-(x), \psi_-(y)
\}
\right)
\left( \frac{1}{2} \dot{\phi} (x) \dot{\phi} (y) -
\frac{1}{2}
\phi^\prime(x) \phi^\prime(y) + \frac{1}{2}U(x) U(y)
\right)
\right.
$$
\begin{equation}
\left. + \frac{1}{2} \left(
\{ \psi_+(x), \psi_+(y) \} + \{ \psi_-(x), \psi_-(y)
\}
\right)
\left(
\frac{1}{2} \phi^\prime(x) U(y) + \frac{1}{2} U(x)
\phi^\prime (y)
\right) \right]
\label{eq:A} \ \ .
\end{equation}

   This is an exact result;  all terms with bosonic commutators cancel.  
The
anticommutators $
\{ \psi_+(x), \psi_-(y) \}
$ and $
\{ \psi_+(y), \psi_-(x) \}
$  all vanish, and also the first line in $\zeta(y)$ vanishes, while  
the second line
would seem to give
$$\zeta(y)=
\int_{-\infty}^{+\infty}  dx \delta (x-y) \left[
\frac{1}{2} \langle \eta(x) \eta(x) \rangle U^{\prime
\prime}
\phi^\prime (x) + \langle \eta^\prime(x)
\eta(x) \rangle U^\prime(x) \right]
$$
\begin{equation} = \int_{-\infty}^{+\infty} dx \delta(x-y)
\frac{\partial}{\partial x}
\left[
\frac{1}{2} \langle
\eta(x) \eta(x) \rangle U^\prime(x) \right] \ \ ;
\end{equation}
\begin{equation}\int_{-\infty}^{\infty}dy\zeta(y) =
\frac{1}{2}
\langle
\eta(x)
\eta(x)
\rangle U^\prime(x) |_{-\infty}^\infty
\label{eq_B} \ \ .
\end{equation} This is the expression obtained in
\cite{rebhan}. Below we show that with appropriate care (i.e., not  
setting $x=y$ too
soon), there is an extra term -- the anomaly. The naive result in  
(\ref{eq_B}) contains a free
field propagator for
$\eta$, because at $x=
\pm
\infty$ the effects of the kink disappear, and, adding the counterterm  
to the central
charge due to mass renormalization, all  quantum corrections to the  
central charge
would seem to vanish. In the  approach of
\cite{shifman}, on the other hand, the central charge contains  a naive  
term
$\phi^\prime U$ and an explicit correction term which is also a total  
derivative and
proportional to
$1/M^2$. Because their $\eta$ propagator contains an extra regulating  
factor
$(k^2+M^2)^{-1}$, the contribution in (\ref{eq_B}) now cancels even  
after
regularization, but because the correction term contains two extra  
derivatives
(to balance the factor $1/M^2$) it yields an extra contribution  
proportional to
$M^2/M^2$, which is the anomaly.

In our case we start from (\ref{eq:A}), but without extra terms as in
\cite{shifman}. We keep $x\ne y$ in (\ref{eq:A}), giving
\begin{equation}
\zeta(y)=\int dx \delta (x-y)[U''(x)\phi_{\rm
kink}'(y)\frac14\langle\eta^2(x)\rangle+
\frac12\langle\eta'(y)\eta(x)\rangle U' (x)+(x \leftrightarrow y)
+\Delta\mu^2 \ {\rm term}] \ \  . \label{eta}
\end{equation} We now show
  that the result is still a total derivative, but instead of the total  
derivative in
(\ref{eq_B}), rather a total derivative with an extra term. The
crucial point is that one cannot replace
$\delta(x-y)\langle\eta'(x)\eta(y)\rangle$ by
$\frac12\delta(x-y)\partial_x\langle\eta^2(x)\rangle$ because
there is a singularity in $\langle\eta'(x)\eta(y)\rangle$ as $x$
tends to $y$.  Setting $\delta(x-y)[\langle \eta'(x)\eta(y)
\rangle -(1/2)\partial_x\langle \eta(x)^2\rangle =0$ would mean
that all terms vanish as in \cite{rebhan}. However, the
singularity which invalidates this equality yields the anomaly
\begin{equation}
\int_{-\infty}^\infty \zeta (x) dx = \left.
\frac{W^{\prime\prime}(\phi)}{4 \pi}
\right|_{-\infty}^{\infty} \ \ ,
\end{equation}
where $W^\prime(\phi) =  U$. Hence
$M^{(1)}=-Z^{(1)}$ in agreement with the invariance of the background  
under $Q_{+}$
(which corresponds to the susy transformation with parameter  
$\epsilon_-$).  Let us
see this explicitly.

The  identity we need is
\begin{equation}
\delta(x-y) \langle \eta^\prime (x) \eta(y)  \rangle ( f(x) - f(y) ) =  
- \frac{1}{2
\pi} \delta(x-y) f^\prime(x) \ \ ,
\label{eq37m}
\end{equation} where $f$ is any smooth function of $x$. The proof of  
this identity
follows from $\langle \eta(x)
\eta(y)
\rangle  = -\frac{1}{2 \pi}
\ln |x-y| + A(x,y)$, where $A$ is a smooth, symmetric function,  
therefore  involving
only nonnegative even powers of $(x-y)$, as can be seen from  
(\ref{m7}-\ref{m9}). The
actual calculation of the anomaly is now very simple.  Expanding all  
contributions in
terms of $x-y$,  and using $\delta(x-y)(x-y)\Delta\mu^2=0$ after  
regularization of
$\Delta\mu^2$,  while
$\delta(x-y)(x-y)\partial_x\langle\eta^2(x)\rangle=0$ does not need   
regularization
because
$\partial_x\langle\eta^2(x)\rangle$ is finite,  we have from
(\ref{eq37m}).
\begin{equation}
\int_{-\infty}^\infty  \zeta(y) dy =
\int_{-\infty}^\infty dxdy
\left[  \frac{1}{4\pi }  \delta(x-y) U^{\prime
\prime}(\phi)
\phi^\prime(x) \right] =  \frac{m}{2 \pi } \ \ .
\end{equation} Here we used $U= \sqrt{\frac{\lambda}{2}}
\left(
\phi^2 -
\frac{\mu^2}{\lambda}
\right) $, $\phi = \frac{\mu}{\sqrt{\lambda}}
\tanh{
\frac{\mu x}{\sqrt{2}}}
$, and $m= \mu \sqrt{2}$.  Again we have the accepted result, and we  
see that point-splitting
regularization yields the same extra term in the central charge as does  
higher-derivative
regularization.

Thus we have shown in a simple way that the term
$\langle \eta^\prime(y) \eta(x) \rangle U^\prime(x)
$ produces the anomaly if one does not set $x=y$ too soon. In  
\cite{shifman} a more
complicated but also more powerful  regularization scheme was used to  
prove this. Our
observation  pinpoints the place where naive methods missed the  
anomaly.  As discussed
extensively  in the  previous sections, one must add to the anomalous  
part the various
contributions to the non-anomalous part of the central charge density.   
This works exactly as in
\cite{shifman}, and of course is completely unaffected by choice of  
regularization  method.

In view of our emphasis on lmr for energy density, it is
reasonable to ask why we do not attempt to apply it to central
charge density.  Looking at (\ref{eta}), one sees that the
expression to be regulated, the bilocal correlator in $\eta(x)$
and $\eta'(y)$, which clearly is not determined by insisting that
the regulated sum ${\rm Im}\langle\eta (x)\dot {\eta }(x)\rangle$
is unchanged between vacuum and kink backgrounds (the entire
content of the lmr prescription). Thus lmr may be applied as a
condition on the expressions in the central charge density, but is
not sufficient to regulate them.

{\it Foundations and conclusions.---}Finally we comment on the physical  
basis for lmr.
In Planck's original formulation of quantum physics \cite{Pl}, the  
number of degrees of freedom is defined by the
available volume in phase space. To fix the total number of modes while  
introducing a
background potential affecting the fluctuations is simply to conserve  
the total phase
space available. The work of Einstein \cite{Ein} and Debye \cite {Deb}  
on crystal vibration contributions to
heat capacity introduced the concept of a local density of degrees of  
freedom,
codifying a notion already found in Boltzmann's lectures on gas theory  
\cite{Bolt}. As was
true for their work, in a lattice approach the number of degrees of  
freedom  per unit
volume evidently does not change when interactions are  introduced, and  
the local mode
density should be equal to this number of degrees of freedom.

Also point splitting methods clarify the meaning  of lmr. Consider the  
bosonic local
mode density regulated by point splitting
\begin{equation}
\rho(k,x)=\int_{-\infty}^{\infty} dy \phi^*\left( k,x-\frac{y}{2}\right)
\phi\left( k,x+\frac{y}{2}\right) f(y) \ \ ,\label{dens}
\end{equation} where $f(y)$ is a function sharply peaked around
$y=0$, with $\int dyf(y)=1$. For large $k$,  the  JWKB approximation for
$\phi(k,x)$ is
\begin{equation}
\phi(k,x)=e^{ikx}e^{-i\int^x dx^{\prime}V(x^{\prime})/2k} \
\ ,
\end{equation} where
$V(x)=U(\phi(x))U^{\prime\prime}(\phi(x)) +
(U^{\prime})^2(\phi(x))-(U^{\prime})^2(\phi(|x|\to\infty))$.  
Substituting this
expression into $\rho(k,x)$ one finds for the integrand of (\ref{dens})
\begin{equation} e^{iky}e^{-i\int_{x-\frac{y}{2}}^{x+\frac{y}{2}}
\frac{V(x^{\prime})}{2k}dx^{\prime}} f(y)
\simeq e^{i\left( k-\frac{V(x)}{2k}\right)y} f(y)
  \ \ .
\end{equation} In the trivial sector
$\rho(k,x)=\tilde{f}(k)$, where $\tilde{f}(k)$ is the Fourier transform  
of
$f(x)$, but in the kink sector one finds a modification
$\rho(k,x)=\tilde{f}\left(k-V(x)/2k\right)$. The energy  density  
therefore contains a
term
$\delta\epsilon(x)=\int\delta\rho(k,x)\frac{1}{2}\omega
\frac{dk}{2\pi}$, and expanding
$\tilde{f}$ we find\footnote{The narrower f(x), the wider the range in $k$ 
contributing to (\ref{ps}), 
so that
in the limit the contribution of any finite range around $k=0$ becomes
negligible. This justifies our use of the JWKB approximation, which is valid for
$k^2\gg m^2,V$.}    

\begin{equation}
\delta\epsilon(x)=  -\int_{-\infty}^{\infty}\frac{dk}{2\pi}
\frac{V(x)}{2k}\frac{\omega}{2}
\frac{\partial}{\partial k}
\tilde{f}(k) =-V(x)/8\pi\ \ . \label{ps}
\end{equation}   For the bosonic kink, this is the anomaly in (\ref{bonomaly}),
 as one may readily check
by direct substitution.

From the JWKB  
form for the wave
function at high energies it follows that the quantity $\Delta
\Lambda (x)$, and hence the anomaly, depends on $x$ only through
the potential felt by the fluctuations.  This in turn implies that
the local anomaly in the energy density is determined at each $x$
by the value of the classical background field $\phi(x)$, as
stated for the central charge density of the supersymmetric case
in \cite{shifman}.

While the above discussion shows that the lmr result for the
bosonic kink follows from point-splitting, it is possible to make
a much stronger statement, that point-splitting implies lmr for a scalar
field in an arbitrary background potential in one space dimension.  The same
JWKB approximation used to determine the shift in effective wave number due
to the potential $V$ can be used also to compute the modulation of the
mode density versus energy (or equivalently, versus asymptotic wavenumber
$k$) in the region of nonzero potential, compared to the asymptotic density
far away. This is a standard calculation, obtaining the wave function and
hence the density to one higher order in $1/k$ than required for the
shift in effective wavenumber.  There is a simple physical mnemonic for the
result of the calculation.  Treating this system as a Schr\"odinger problem 
with
``Hamiltonian'' $\omega^2$, the modulating factor is simply the ratio of the
asymptotic classical velocity to the local velocity:
\begin{equation}
\rho(x)/\rho(\infty)=v/v(x)=k/\sqrt{k^2-V(x)} \ \ ,
\end{equation} 
or
\begin{equation}
\delta\rho /\rho\sim V(x)/2k^2 \ \ .
\end{equation}
Integrating the above expression from a nominal sharp cutoff $\Lambda$ to
$\infty$, we see that to have the same integrated mode number density above
the cutoff in the presence of a potential
$V(x)$ as in a trivial background, we must shift the cutoff by
\begin{equation}
\delta \Lambda = V(x)/2\Lambda \ \  ,
\end{equation}
exactly the amount implied by point-splitting as found in the
discussion leading to (\ref{ps}) above.  

The equality between the wave number shift $\delta k(x)$ and the integral
over the density shift is reminiscent of an unsubtracted dispersion
relation.  Possibly the implementation of lmr in higher dimensions would
require the equivalent of subtracted dispersion relations to compensate
for the increasing divergence of energy density with cutoff.

We have seen that point-splitting, a regularization scheme which
is local but not necessarily useful beyond one-loop order, gives
the same anomaly in the central charge found in \cite{shifman}
with higher-derivative regularization, and also implies lmr for
the bosonic energy density with arbitrary background potential. As explained
above, the converse is not true: lmr does not contain the full content of
point-splitting regularization.  Nevertheless, it is appealing that it
captures in one line the above sequence of equations required to obtain the
anomaly from point-splitting regularization.  Thus, as stated
already in the introduction, lmr is a simple, easily-used tool for
a special but extremely important application, the computation of
local one-loop energy densities.

For the supersymmetric case, one gains insight into the requirement of  
equal bosonic
and fermionic mode densities by considering the $N=2$ theory, where  
there is an abelian
charge density which should be invariant under supersymmetry.  That  
requirement
automatically imposes the constraint represented by lmr.

While all of the above are appealing arguments, the accepted criterion  
for determining
the validity of a regulation procedure is to insert regulators into the  
action in such
a way that all relevant symmetries are satisfied at the regulator  
level, and then to
deduce consequences for specific quantities.  Thus in the present case  
a definitive
check on the validity of lmr would be to use, for example,  
higher-derivative
regulation (which obeys supersymmetry), and check that  this scheme  
implies lmr.  This
important analysis remains to be done.  As an alternative, one might be able to
 prove 
that point-splitting preserves susy in
models with solitons, and then extend our deduction of lmr from point-splitting
in the bosonic case to the susy case.  This is something to which we intend to 
return in the future.

To summarize, lmr permits one to isolate and then compute directly
the anomalous contribution to the energy density of the bosonic or
susy kink.  Expressed most conservatively, lmr at the least gives
a simple interpretation of the anomaly as the shift in energy
density required to equalize mode densities.  In fact as we have just seen,
at least for Bose fields in one space dimension with arbitrary scalar
background potential, lmr follows from point-splitting regularization of the
energy density.  In
addition we found remarkable phase space factorization identities for the
non-anomalous   contributions to
the energy density, which might hold for all reflectionless potentials.  
  These
non-anomalous contributions are independent of the regularization  
method (though
sensitive to renormalization conditions because they only are  
convergent after
subtraction of the mass counter term).   Elsewhere
\cite{future} we compute the divergent energy density at the boundary  
of the kink with
supersymmetric boundary conditions, and obtain an analytic expression  
for the anomaly
near the boundary, which in the limit when the regulator energy goes to  
infinity
becomes a delta-function contribution just at the boundary, in  
agreement with
expectations of \cite{shifman}.    It would be interesting to explore  
local
mode regularization further, comparing with complete regularization schemes
and   studying
solitons in higher dimensions such as the magnetic monopole, and also  
explore phase
space factorization, seeking a theoretical basis as well as additional  
examples.

We thank Kevin Cahill, Ludwig Faddeev, and Stefan  Vandoren for
discussions, and in particular Arkady Vainshtein for useful
detailed comments on a number of occasions. We note an independent
proposal of lmr by Robert Wimmer \cite{wim}, which appeared after
submission of the present work.

\end{document}